\documentclass[twocolumn,superscriptaddress]{revtex4-2}

\usepackage{graphicx}
\usepackage{dcolumn}
\usepackage{bm}

\usepackage{color}
\usepackage{amsmath}
\usepackage{float}
\usepackage{booktabs}
\usepackage{caption}
\usepackage{ragged2e}  
\DeclareUnicodeCharacter{2550}{\ensuremath{\equiv}}
\usepackage{hyperref}
\definecolor{darkred}{rgb}{0.25,0,0}
\definecolor{darkgreen}{rgb}{0,0.25,0}
\definecolor{darkblue}{rgb}{0,0,1}
\hypersetup{colorlinks,linkcolor=darkblue,filecolor=black,urlcolor=darkblue,citecolor=darkblue}
\usepackage{multirow}
\usepackage{textcomp}
\usepackage[compact]{titlesec}
\vspace{-0.5cm} 
\DeclareUnicodeCharacter{0393}{-}
\usepackage{ragged2e}
\begin{document}

\preprint{APS/123-QED}

\title{Diverse Responses in Lattice Thermal Conductivity of $n$-type/$p$-type Semiconductors Driven by Asymmetric Electron-Phonon Interactions}


\author{Jianshi Sun}
\affiliation{Institute of Micro/Nano Electromechanical System and Integrated Circuit, College of Mechanical Engineering, Donghua University, Shanghai 201620, China
}%

\author{Shouhang Li}
\email{shouhang.li@dhu.edu.cn}
\affiliation{Institute of Micro/Nano Electromechanical System and Integrated Circuit, College of Mechanical Engineering, Donghua University, Shanghai 201620, China
}%

\author{Zhen Tong}
\affiliation{
School of Advanced Energy, Sun Yat-Sen University, Shenzhen 518107, China
}

\author{Cheng Shao}
\affiliation{
 Thermal Science Research Center, Shandong Institute of Advanced Technology, Jinan, Shandong 250103, China
}

\author{Han Xie}
\affiliation{
Artificial Intelligence Lab, SAIC Motor Corporation, Shanghai 200433, China 
}
\affiliation{School of Energy and Materials, Shanghai Polytechnic University, Shanghai 201209, China}

\author{Meng An}
\affiliation{
 Department of Mechanical Engineering, The University of Tokyo, 7-3-1 Hongo, Bunkyo, Tokyo, 113-8656, Japan
}

\author{Chuang Zhang}
\affiliation{Institute of Micro/Nano Electromechanical System and Integrated Circuit, College of Mechanical Engineering, Donghua University, Shanghai 201620, China
}%

\author{Xiongfei Zhu}
\affiliation{Institute of Micro/Nano Electromechanical System and Integrated Circuit, College of Mechanical Engineering, Donghua University, Shanghai 201620, China
}%

\author{Chen Huang}
\affiliation{Institute of Micro/Nano Electromechanical System and Integrated Circuit, College of Mechanical Engineering, Donghua University, Shanghai 201620, China
}%

\author{Yucheng Xiong}
\affiliation{Institute of Micro/Nano Electromechanical System and Integrated Circuit, College of Mechanical Engineering, Donghua University, Shanghai 201620, China
}%

\author{Xiangjun Liu}
\email{xjliu@dhu.edu.cn}
\affiliation{%
 Institute of Micro/Nano Electromechanical System and Integrated Circuit, College of Mechanical Engineering, Donghua University, Shanghai 201620, China
}%

\date{\today}

\begin{abstract}
Accurately assessing the impact of electron-phonon interaction (EPI) on the lattice thermal conductivity of semiconductors is crucial for the thermal management of electronic devices and a unified physical understanding of this issue is highly desired. In this work, we predict the lattice thermal conductivities of typical direct and indirect bandgap semiconductors accounting for EPI based on mode-level first-principles calculations. It is found that EPI has a larger effect on the lattice thermal conductivity of $p$-type doping compared to $n$-type doping in the same semiconductor at high charge carrier concentrations. The stronger EPI in $p$-type doping is attributed to the relatively higher electron density of states caused by the relatively larger $p$-orbital component. Furthermore, EPI has a stronger influence on the lattice thermal conductivity of $n$-type indirect bandgap semiconductors than $n$-type direct bandgap semiconductors. This is attributed to the relatively lower electron density of states in direct bandgap semiconductors stemming from the $s$-orbital component. This work reveals that there exist diverse responses in lattice thermal conductivity of $n$-type/$p$-type semiconductors, which can be attributed to asymmetric EPIs.
\end{abstract}

\maketitle
\section{INTRODUCTION}
Semiconductors serve as the cornerstone of the modern information technology industry. Charge carrier transport in semiconductors is responsible for electrical current, yet their scattering with the lattice is inevitable. The electron-phonon interaction (EPI) has been extensively studied for its impacts on electrical transport properties, including carrier mobility, electrical thermal conductivity, and superconductivity\cite{ziman2001electrons,ponce2019hole,giustino2017electron,ma2020electron,liu2017first,xia2021limits,li2022anomalously,li2021thermal,li2020anomalous,ponce2019route}. However, the EPI can also affect the thermal transport properties of thermoelectric materials\cite{zhang2013high} and nano-electronic devices\cite{del2011nanometre} when the charge carrier concentrations reach as high as $10^{20}$ to $10^{21} \, \text{cm}^{-3}$. Efficient heat dissipation is crucial for the reliability and stability of semiconductor-based devices\cite{warzoha2021applications}.

To accurately obtain the thermal transport properties of semiconductors, the methodology of extracting force constants from first-principles calculations combined with the Peierls-Boltzmann transport equation (PBTE) has been demonstrated to robustly predict the lattice thermal conductivity of intrinsic semiconductors without requiring any parameters\cite{lindsay2013first,li2013thermal,feng2017four,lindsay2013ab,sun2023light,gu2020thermal,feng2014prediction}. In addition, PBTE can provide abundant physical information, such as the phonon mean free path and Grüneisen parameter, which are critical to understanding the thermal transport mechanisms. The key step in iteratively solving the PBTE is to obtain various phonon scattering rates. Milestone progress has been achieved in the solution of phonon-phonon, phonon-isotope, phonon-impurity, and phonon-boundary scattering rates\cite{lindsay2016first}. To achieve excellent electrical transport properties in semiconductor-based devices, doping is typically required to attain higher carrier concentrations. For instance, the electrical conductivity of $n$-type AlN increases by four orders of magnitude as the silicon doping concentration ranges from \(1.2 \times 10^{21} \, \text{ to } \, 2.5 \times 10^{21} \, \text{cm}^{-3}\)\cite{hermann2005highly}. Similarly, the resistivity is reduced to \(2.7 \times 10^{-2} \, \Omega \cdot \text{cm}\) by increasing the concentration of implanted aluminum ion to \(8 \times 10^{20} \, \text{cm}^{-3}\) for $p$-type 4H-SiC\cite{nipoti2013conventional}. However, high carrier concentration inevitably introduces an additional phonon-electron scattering term, which has long been ignored in previous empirical models for lattice thermal conductivity\cite{tritt2005thermal}.

Liao \textit{et al}. found that the lattice thermal conductivity of silicon is significantly reduced by $\sim37$\% for an electron concentration of $10^{21} \, \text{cm}^{-3}$ and by $\sim45$\% for a hole concentration of $10^{21} \, \text{cm}^{-3}$ at room temperature\cite{liao2015significant}. Subsequent experiments using ultrafast photoacoustic spectroscopy confirmed significant suppression of phonon propagation by phonon-electron scattering in silicon at room temperature\cite{liao2016photo}. Similar trends were also observed in other semiconductors such as 3C-SiC\cite{wang2017strong} and SiGe alloy\cite{xu2019effect}. In the aforementioned semiconductors, both electrons and holes are strongly coupled with phonons, resulting in a substantial decrease in lattice thermal conductivity. However, our previous work shows that EPI has a weak impact on the lattice thermal conductivity of $n$-type GaN at ultra-high electron concentrations, attributed to fewer phonon-electron scattering channels\cite{sun2024weak}. In contrast, there is a significant reduction in the lattice thermal conductivity of $p$-type GaN at high hole concentrations. Therefore, there is a lack of a unified physical understanding regarding the impact of EPI on the lattice thermal conductivities of $n$-type and $p$-type semiconductors. It should be also noted that the computation of phonon-electron scattering rates is extremely expensive and laborious. Hence, revealing the physical mechanism has a great promotion in the prior assessment of the EPI effects on the lattice thermal conductivity of semiconductors.

In this work, we investigate the EPI effects on the lattice thermal conductivity of typical direct and indirect bandgap compound semiconductors with $n$-type and $p$-type doping, using rigorous mode-level first-principles calculations combined with the PBTE. For direct bandgap semiconductors, wurtzite structure GaN and AlN are chosen, while for the indirect bandgap semiconductor, wurtzite structure 2H-SiC is chosen. We provide a comprehensive analysis of the contribution terms attributed to EPI, including electron density of state (DOS), Fermi surface nesting function, electron-phonon matrix element. We reveal that the diverse responses in lattice thermal conductivity of $n$-type/$p$-type semiconductors stem from asymmetric EPIs.

\section{THEORY AND METHODS}
Combining the linearized phonon Boltzmann transport equation and Fourier's law, the lattice thermal conductivity (${\kappa}_{\text{lat}}$) can be calculated as\cite{broido2007intrinsic}
\begin{equation}
\begin{split}
    \kappa_{\text{lat}, \alpha \beta} &= \sum_{\lambda} c_{v, \lambda} v_{\lambda, \alpha} v_{\lambda, \beta} \tau_{\lambda},
\end{split}
\label{SE1}
\end{equation}
where $\alpha$ and $\beta$ are the Cartesian coordinates,  $\lambda \equiv(\mathbf{q}, v)$ denotes the phonon mode with wave vector $\mathbf{q}$ and phonon polarization $v$. The phonon specific heat capacity is denoted as $c_{\mathrm{v}, \lambda}=\frac{1}{V} \hbar \omega_{\lambda} \frac{\partial n_{\lambda}}{\partial T}$, where ${V}$ is the volume of the primitive cell, $\omega_{\lambda}$ is the phonon frequency, $n_{\lambda}$ is the Bose-Einstein distribution at temperature $T$. $v_{\lambda}$ is the phonon group velocity, and $\tau_{\lambda}$ is the phonon relaxation time, which can be obtained using Matthiessen’s rule as $1/\tau_{\lambda}$=$1/\tau_{\lambda}^{\mathrm{ph}-\mathrm{ph}}$+$1/\tau_{\lambda}^{\mathrm{ph}-\text{iso}}$+$1/\tau_{\lambda}^{\mathrm{ph}-\text{el}}$, where $1/\tau_{\lambda}^{\mathrm{ph}-\mathrm{ph}}$ is the phonon-phonon scattering rate, $1/\tau_{\lambda}^{\mathrm{ph}-\text{iso}}$ is the phonon-isotope scattering rate, and $1/\tau_{\lambda}^{\mathrm{ph}-\text{el}}$ is the phonon-electron scattering rate.

The phonon-phonon scattering rate due to three-phonon scattering is given by Fermi's golden rule as\cite{albers1976normal}
\begin{equation}
\begin{aligned}
    \frac{1}{\tau_{\lambda}^{\mathrm{ph}-\mathrm{ph}}} &= 2 \pi \sum_{\lambda_{1} \lambda_{2}}\left|V_{\lambda \lambda_{1} \lambda_{2}}\right|^{2} \\
    &\times \left[\frac{1}{2}\left(1+n_{\lambda_{1}}^{0}+n_{\lambda_{2}}^{0}\right) \delta\left(\omega_{\lambda}-\omega_{\lambda_{1}}-\omega_{\lambda_{2}}\right) \right. \\
    &\quad + \left. \left(n_{\lambda_{1}}^{0}-n_{\lambda_{2}}^{0}\right) \delta\left(\omega_{\lambda}+\omega_{\lambda_{1}}-\omega_{\lambda_{2}}\right)\right],
\end{aligned}
\label{SE2}
\end{equation}
$V_{\lambda \lambda_{1} \lambda_{2}}$ denote the three-phonon scattering matrix element, which is related to the third-order force constants. $\delta$ is the Dirac delta function which ensures the conservation of energy during the scattering processes. Recent studies have shown that four-phonon scattering has weak effects on the lattice thermal conductivity of GaN, 2H-SiC, and AlN at room temperature\cite{sun2024weak,yang2019stronger,zhang2021gpu_pbte,srivastava2009anharmonic}. Consequently, four-phonon scattering is not incorporated into our calculations.

The phonon-isotope scattering rate can be calculated based on the Tamura theory as\cite{tamura1983isotope}
\begin{equation}
\begin{aligned}
    \frac{1}{\tau_{\lambda}^{\mathrm{iso}}}=\frac{\pi \omega_{\lambda}^{2}}{2 N} \sum_{i \in u . c .} g(i)\left|\boldsymbol{e}_{\lambda^{\prime}}^{*} \cdot \boldsymbol{e}_{\lambda}(i)\right|^{2} \delta\left(\omega_{\lambda}-\omega_{\lambda^{\prime}}\right),
\end{aligned}
\label{SE3}
\end{equation}
where $\boldsymbol{e}$ is the normalized eigenvector of phonon mode $\lambda$ and the asterisk denotes the complex conjugate, $g(i)=\sum_{i} n_{i}(j)\left[1-m_{i}(j) / \bar{m}_{i}(j)\right]^{2}$, where $n_{i}(j)$ and $m_{i}(j)$ are the concentration and atomic mass of the \textit{i}th substitution atom, respectively. $\bar{m}_{i}(j)$ is the average mass of the \textit{j}th atom in the unit cell.

The phonon-electron scattering rate is related to the imaginary part of the phonon self-energy, which can be expressed as\cite{ponce2016epw}
\begin{equation}
\begin{aligned}
    \frac{1}{\tau_{\lambda}^{\text {ph-el}}} &= -\frac{2 \pi}{\hbar} \sum_{m n, \mathbf{k}}\left|g_{m n}^{v}(\mathbf{k}, \mathbf{q})\right|^{2} \\
    &\quad \times \left(f_{n \mathbf{k}}-f_{m \mathbf{k}+\mathbf{q}}\right) \delta\left(\varepsilon_{m \mathbf{k}+\mathbf{q}}-\varepsilon_{n \mathbf{q}}-\hbar \omega_{\lambda}\right),
\end{aligned}
\label{SE4}
\end{equation}
where $g_{m n, v}(\mathbf{k}, \mathbf{q})$ is the electron-phonon matrix element, which quantifies probability amplitude for scattering between the electronic state $n \mathbf{k}$ and $m \mathbf{k}+\mathbf{q}$. The Fermi-Dirac distribution funciton is denoted as $f(\varepsilon)=\frac{1}{\exp \left((\varepsilon-\varepsilon_{F}) / k_{\mathrm{B}} T\right)+1}$, where $k_{\mathrm{B}}$ is the Boltzmann constant, $\varepsilon$ is the electron energy, and $\varepsilon_{F}$ is the Fermi energy.

\begin{figure*}[hbpt]
    \includegraphics[width=2\columnwidth]{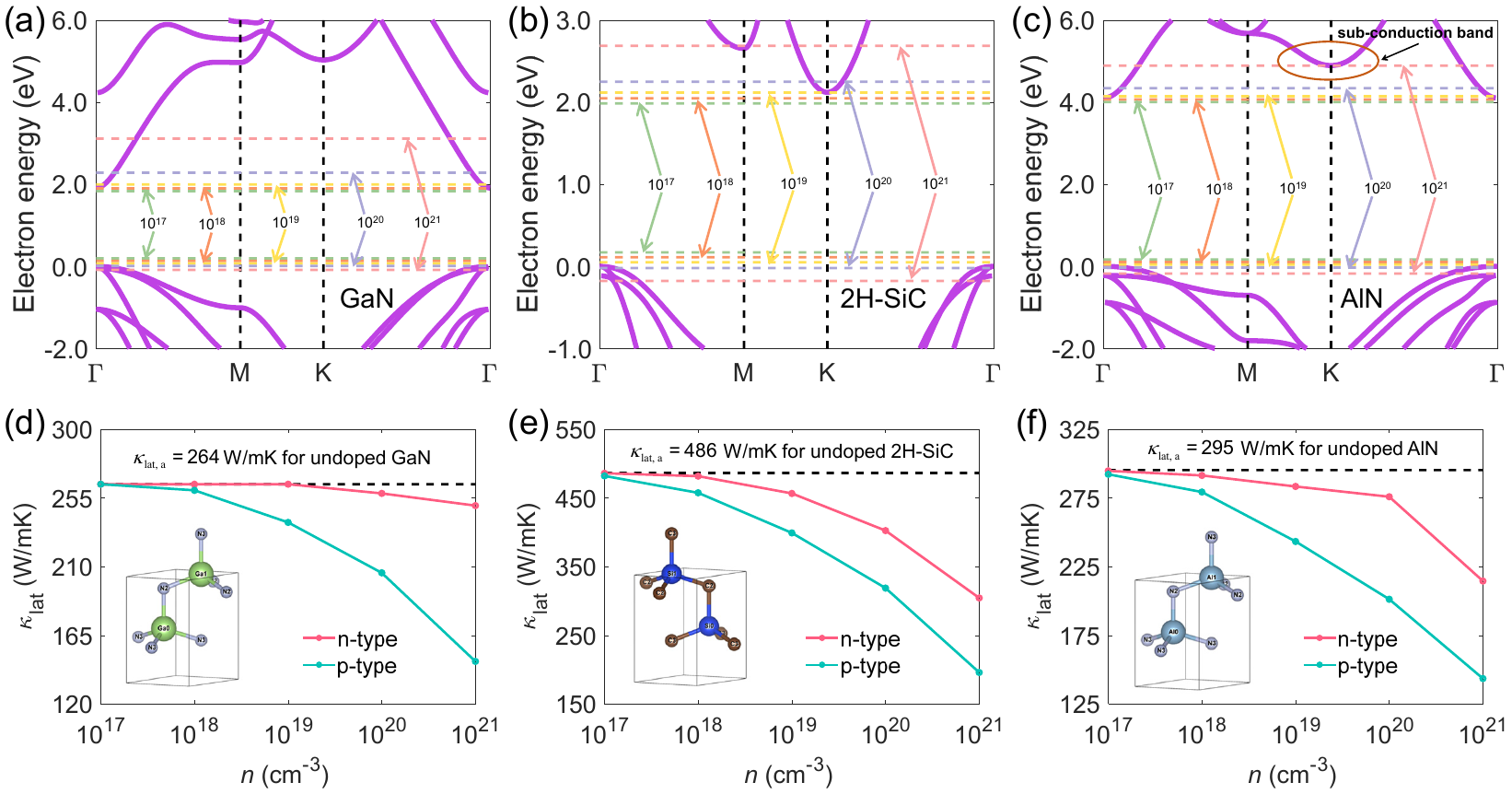}
    \caption{\justifying Band structures of (a) GaN, (b) 2H-SiC, and (c) AlN along the high symmetry paths, respectively. The horizontal lines are Fermi energy related to the carrier concentrations of $10^{17}$ (green), $10^{18}$ (orange), $10^{19}$ (yellow), $10^{20}$ (light purple), and $10^{21}$ (light pink) $\text{cm}^{-3}$ at room temperature. The electron energy is normalized to the VBM. The $\kappa_{1\text{at},a}$ as a function of carrier concentration at room temperature for undoped, $n$-type, and $p$-type (d) GaN, (e) 2H-SiC, and (f) AlN, respectively. The crystal structure schematic illustration for different semiconductors are inserted to the bottom of the (d-f).}
    \label{fig: Figure 1}
\end{figure*}

The first-principles calculations are performed using the Quantum Espresso package\cite{giannozzi2009quantum}. The electron exchange-correlation functional is treated by the generalized gradient approximation of Perdew-Burke-Ernzerhof (PBE)\cite{perdew2008restoring} and optimized full relativistic norm-conserving pseudopotentials\cite{hamann2013optimized} from PseudoDojo\cite{van2018pseudodojo}. 
The lattice vectors and atomic positions are fully relaxed based on the Broyden-Fretcher-Goldfarb-Shanno (BFGS) optimization method\cite{fletcher1970new,broyden1970convergence,goldfarb1970family,shanno1970conditioning} and the convergence thresholds for energy and force are set to $10^{-8}$ Ry and $10^{-8}$ Ry/Bohr. The kinetic energy cutoff for plane waves is set to be 120 Ry. The Brillouin zone is sampled with a 12×12×8 Monkhorst-Pack $\mathbf{k}$-point mesh to ensure convergence for self-consistent calculation. The threshold of electron total energy is set to be $10^{-10}$ Ry. The lattice constants of GaN, AlN, and 2H-SiC agree well with the experimental data\cite{schulz1977crystal,schulz1979structure}, as shown in [Table S1 Supporting Material]. The harmonic force constants are calculated from density-functional perturbation theory (DFPT)\cite{fugallo2013ab,baroni2001phonons}. The $\mathbf{q}$-point mesh is set to be 6×6×4 and the energy threshold is $10^{-17}$ Ry to guarantee convergence. Also, the dielectric constant and Born effective charge are calculated to account for the long-range electrostatic interactions. The phonon dispersions of GaN, AlN, and 2H-SiC are in good agreement with the experimental data\cite{ruf2001phonon,schwoerer1999phonons}, as shown in [Fig. S1, Supporting Material]. The cubic force constants are also calculated from DFPT\cite{paulatto2013anharmonic}. The $\mathbf{q}$-point mesh is set to be 3×3×2. The phonon-electron scattering rates are calculated by our in-house modified Electron-Phonon Wannier Package\cite{ponce2016epw}. The electron band structures calculated by density functional theory agree quite well with that obtained through the Wannier technique, as shown in [Fig. S2, Supporting Material]. The electron-phonon coupling matrix elements are first calculated under coarse $\mathbf{k/q}$-point meshes (12×12×8/6×6×4), and then interpolated to dense $\mathbf{k/q}$-point meshes (36×36×24/21×21×14) with Wannier interpolation technique\cite{marzari2012maximally}. The convergence of the phonon-electron scattering rates with respect to the $\mathbf{k}$-point mesh is checked in [Fig. S3, Supporting Material]. The in-house modified D3Q package\cite{paulatto2013anharmonic} is employed to calculate the ${\kappa}_{\text{lat}}$, incorporating the phonon-electron scattering rate using the iterative calculation scheme\cite{sun2024weak,li2022anomalously,li2020anomalous}. The phonon-isotope scattering is incorporated in all calculations. ${\kappa}_{\text{lat}}$ is converged with respect to a $\mathbf{q}$-point mesh of 21×21×14. The rigid shift of the Fermi energy is utilized to imitate the change of carrier concentration\cite{li2019resolving}.

\section{RESULTS AND DISCUSSIONS}
Figs. \ref{fig: Figure 1}(a-c) show the electron band structures of GaN, 2H-SiC, and AlN along the high-symmetry path $\Gamma$-M-K-$\Gamma$ in the first Brillouin zone. The impact of spin-orbit coupling (SOC) on ${\kappa}_{\text{lat}}$ is considered in this work\cite{sun2024weak}. The Fermi energy corresponding to different carrier concentrations is depicted by horizontal dashed lines. The band structures of GaN and AlN show that both the valence band maximum (VBM) and conduction band minimum (CBM) are located at the $\Gamma$-point, resulting in direct bandgaps of 1.86 and 4.11 eV, respectively. For the band structure of 2H-SiC, the VBM is located at the $\Gamma$-point and the CBM is located at the K-point, leading to an indirect bandgap of 2.12 eV. All bandgaps are significantly underestimated compared to experimental data\cite{dingle1971absorption,harris1995properties,perry1978optical,monemar1974fundamental}, attributed to the well-known limitations of DFT\cite{chan2010efficient}. Nevertheless, this discrepancy does not impact our conclusions regarding ${\kappa}_{\text{lat}}$ since only electron modes near the band edges contribute to EPI, and the profiles of our DFT band structures closely align with experimental data\cite{dingle1971absorption,harris1995properties,perry1978optical,monemar1974fundamental}. Although GaN and AlN are both direct bandgap semiconductors, the sub-conduction band of AlN is quite close to its CBM, as shown in Fig. \ref{fig: Figure 1}(c). This phenomenon results in an anomalous decrease in   when the electron concentration reaches $10^{21} \, \text{cm}^{-3}$, as will be discussed later.

\begin{figure*}[hbpt]
    \centering
    \includegraphics[width=2.00\columnwidth]{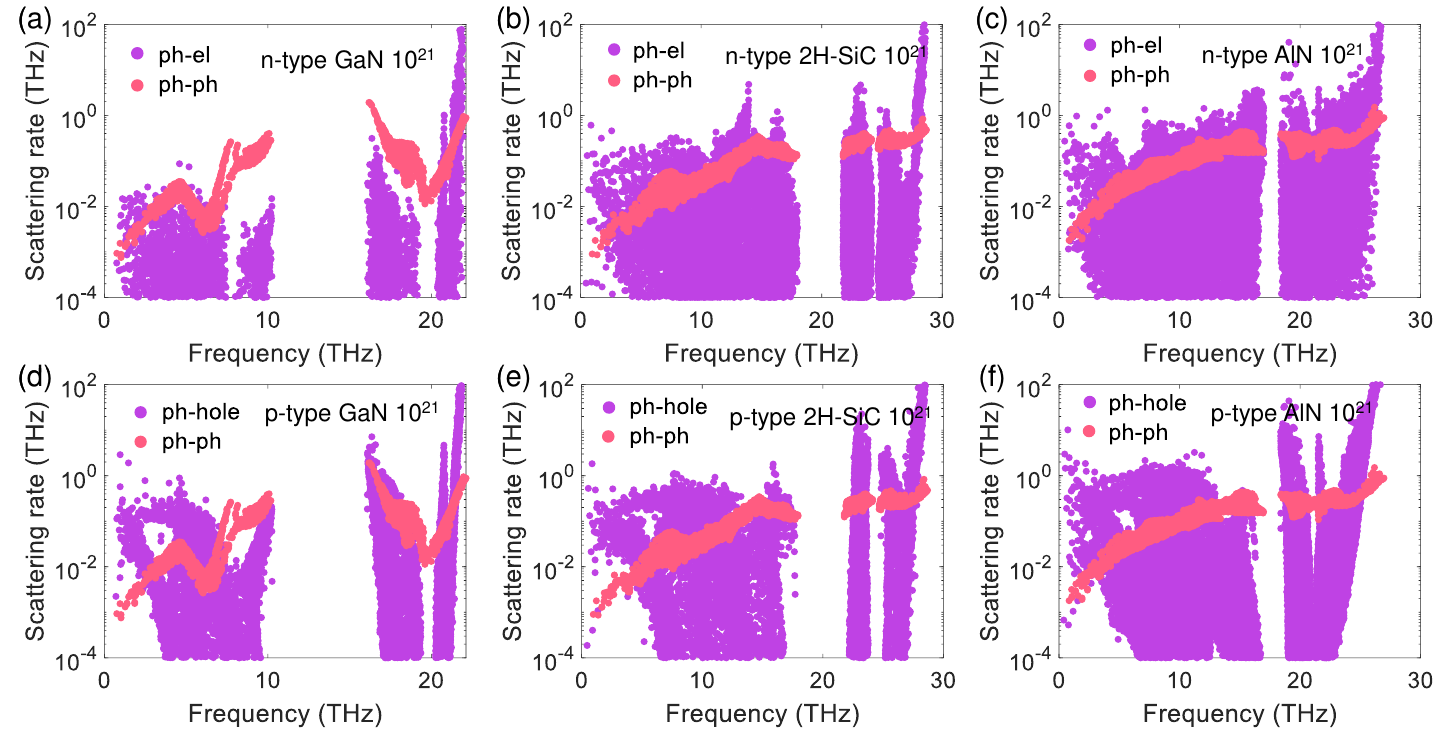}
    \caption{\justifying Phonon-phonon scattering rates, (a-c) phonon-electron scattering rates, and (d-f) phonon-hole scattering rates for GaN, AlN, and 2H-SiC at room temperature with the carrier concentration of $10^{21} \, \text{cm}^{-3}$.}
    \label{fig: Figure 2}
\end{figure*}

Figs. \ref{fig: Figure 1}(d-f) show the in-plane lattice thermal conductivity ($\kappa_{1\text{at},a}$) as a function of carrier concentrations. The $\kappa_{1\text{at},a}$ at room temperature for undoped GaN, 2H-SiC, and AlN is 264 W/mK, 486 W/mK, and 295 W/mK, respectively. Our calculated $\kappa_{1\text{at},a}$ agrees well with the experimental data\cite{jezowski2015thermal,cheng2020experimental} and former theoretical results\cite{protik2017phonon}. The quantitative differences between our work and experimental values may be attributed to impurities, defects, or experimental sample quality, while the slight disparities between our work and theoretical values may stem from the different pseudopotentials used in DFT calculations. With phonon-electron scattering further included, the $\kappa_{1\text{at},a}$ of $n$-type GaN falls within the range of 250-264 W/mK, which is quite close to the undoped case. In contrast, $\kappa_{1\text{at},a}$ dramatically decreases from 264 W/mK to 148 W/mK within the hole concentration range of $10^{17} \, \text{ to } \, 10^{21} \, \text{cm}^{-3}$. In both $n$-type and $p$-type 2H-SiC, there is a notable decrease in $\kappa_{1\text{at},a}$, with values dropping from 486 W/mK to 305 W/mK and from 482 W/mK to 196 W/mK, respectively. Interestingly, the variation in $\kappa_{1\text{at},a}$ exhibits an anomalous trend for $n$-type AlN. When the electron concentration locates in the range of $10^{17} \, \text{ to } \, 10^{20} \, \text{cm}^{-3}$, $\kappa_{1\text{at},a}$ experiences a slight decrease. As the electron concentration rises from $10^{20} \, \text{ to } \, 10^{21} \, \text{cm}^{-3}$, there is an abrupt 22\% decrease in $\kappa_{1\text{at},a}$. Overall, we observe two characteristics of the impact of EPI on $\kappa_{1\text{at},a}$: (1) Hole doping has stronger effects on $\kappa_{1\text{at},a}$ compared to electron doping. (2) The effects of electron doping on $\kappa_{1\text{at},a}$ is stronger in indirect bandgap semiconductors compared to direct bandgap semiconductors, except for the case of AlN with the electron concertation of $10^{21} \, \text{cm}^{-3}$. 

To reveal the underlying mechanisms, the mode-level phonon-electron and phonon-hole scattering rates for GaN, 2H-SiC, and AlN at the carrier concentration of $10^{21} \, \text{cm}^{-3}$ are calculated, as shown in Fig. \ref{fig: Figure 2}. In the following discussions, we mainly focus on the phonon-electron and phonon-hole scattering rates below the phonon frequency gap since they primarily contribute to ${\kappa}_{\text{lat}}$. The phonon-electron and phonon-hole scattering rates for high-frequency optical phonons are relatively larger due to the Fröhlich interaction in polar materials\cite{sheih1995electron}. However, the contribution of high-frequency optical phonon modes to the ${\kappa}_{\text{lat}}$ of GaN, 2H-SiC, and AlN can be neglected\cite{garg2018spectral,sun2024weak,protik2017phonon}. For GaN, the phonon-electron scattering rate is significantly lower than the phonon-phonon scattering rate, whereas the phonon-hole scattering rate is notably higher than the phonon-phonon scattering rate. However, both phonon-electron and phonon-hole scattering rates in 2H-SiC are significantly higher than the phonon-phonon scattering rate, and the latter is quantitatively larger than the former. However, a similar trend is also observed in the direct bandgap semiconductor AlN at the carrier concentration of $10^{21} \, \text{cm}^{-3}$. In contrast, the trend in AlN is similar to that of GaN when the carrier concentration is $10^{20} \, \text{cm}^{-3}$, as shown in [Fig. S4, Supporting Material].

\begin{figure*}[hbpt]
    \centering
    \includegraphics[width=2.00\columnwidth]{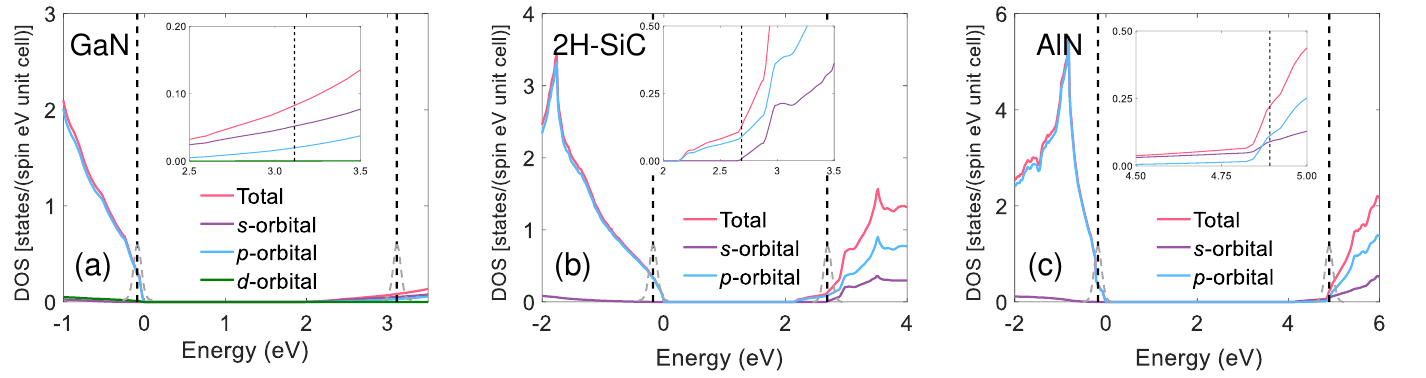}
    \caption{\justifying (a) Total and partial DOS near the valence- and conduction-band edges of for (a) GaN, (b) 2H-SiC, and (c) AlN, respectively. The gray dotted curve represents the Fermi window, indicating the relevant range of energies. The electron energy is normalized to the VBM. The position of the Fermi energy for electron and hole concentrations of $10^{21} \, \text{cm}^{-3}$ is indicated with black dash lines. The enlarged pictures of total and partial DOS for Fermi energy with the electron concentration of $10^{21} \, \text{cm}^{-3}$ are inserted above in (a-c).}
    \label{fig: Figure 3}
\end{figure*}

The differences in the phonon-electron and phonon-hole scattering rates at the same concentration are mainly due to variations in the electron DOS within the Fermi window. Fig. \ref{fig: Figure 3} shows that the DOS for holes is significantly larger than that of electrons at the carrier concentration of $10^{21} \, \text{cm}^{-3}$, exhibiting a pronounced asymmetric profile. For typical semiconductors, the states near the band edges exhibit behavior similar to the $s$ orbital and the three $p$ orbitals observed in individual atoms\cite{vasileska2017computational}. Given that the $s$ orbital predominantly contributes to the DOS, the electron band is more dispersive, resulting in a lower DOS. Conversely, the dominance of the $p$-orbital leads to minimal band dispersion, thereby resulting in a higher DOS\cite{walsh2008origins,yoodee1984effects,reynolds1999valence}. It is found that the $p$-orbital components near the Fermi level with the hole concentration of $10^{21} \, \text{cm}^{-3}$ are significantly higher compared to those for the electron concentration of $10^{21} \, \text{cm}^{-3}$ for all semiconductors studied in this work, as shown in Fig. \ref{fig: Figure 3}. Therefore, EPI has weak effects on the ${\kappa}_{\text{lat}}$ of $n$-type semiconductors, while they have strong effects on that of $p$-type semiconductors. 

On the other hand, the conduction band component at the $\Gamma$-point of direct bandgap semiconductors is an $s$ orbital, while the valence band component is a linear combination of $p$ orbitals. However, there is some amount of $p$ orbital mixed into $s$ orbital in the conduction band of indirect bandgap semiconductors\cite{vasileska2017computational}. It should be noted that as semiconductors become more indirect, the contribution of $p$ orbital components increases. That’s why the $p$-orbital components of 2H-SiC are higher than those of GaN as shown in the inset picture in Figs. \ref{fig: Figure 3}(a) and (b). As a result, the impact of electron doping on ${\kappa}_{\text{lat}}$ is stronger in 2H-SiC compared to GaN. For AlN, the DOS between the CBM and sub-conduction band is primarily dominated by $s$ orbital, similar to the case in direct bandgap semiconductors. However, the DOS is primarily dominated by $p$ orbital above the sub-conduction band (4.77 eV), leading to a significant increase in DOS [inset picture in Fig. \ref{fig: Figure 3}(c)]. Therefore, there is a noticeable decrease in ${\kappa}_{\text{lat}}$ of $n$-type AlN at the concentration of $10^{21} \, \text{cm}^{-3}$.

\begin{figure*}[hbpt]
    \centering
    \includegraphics[width=2.00\columnwidth]{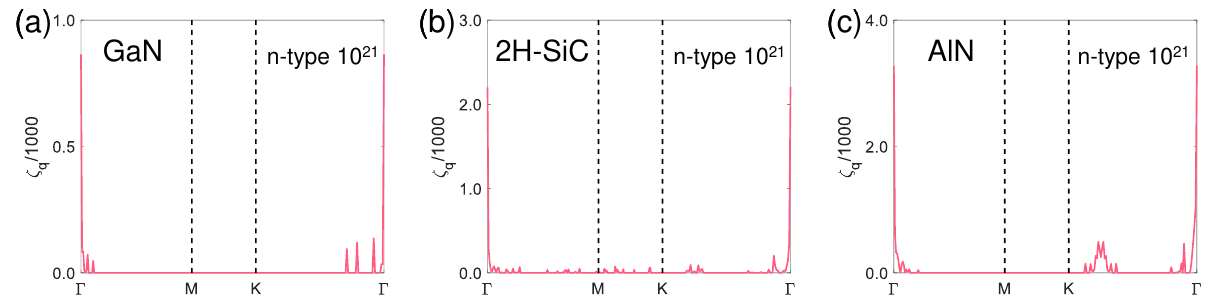}
    \caption{\justifying The Fermi surface nesting functions for (a) GaN, (b) 2H-SiC, and (c) AlN at room temperature, respectively. The carrier concentration is $10^{21} \, \text{cm}^{-3}$ in all cases.}
    \label{fig: Figure 4}
\end{figure*}

According to Eq. \ref{SE4} and its variant derived from the double delta approximation\cite{li2021thermal,li2020anomalous,wang2016first}, the electron DOS largely determines the Fermi surface nesting function ($\zeta_{\mathbf{q}}$), which can be expressed as
\begin{equation}
    \zeta_{\mathbf{q}}=\sum \delta\left(\varepsilon_{n \mathbf{k}}-\varepsilon_{F}\right) \delta\left(\varepsilon_{m \mathbf{k}+\mathbf{q}}-\varepsilon_{F}\right),
    \label{SE5}
\end{equation}
$\zeta_{\mathbf{q}}$ quantifies the phonon-electron scattering channels\cite{li2018fermi}. As shown in Fig. \ref{fig: Figure 4}, $\zeta_{\mathbf{q}}$ of $n$-type GaN is much lower than that of $n$-type 2H-SiC, which results in its relatively lower phonon-electron scattering rates. Note that there is an additional scattering channel generated by the sub-conduction band located at the K-point, which is 0.73 eV away from the CBM in AlN as the electron concentration increases from $10^{20} \, \text{ to } \, 10^{21} \, \text{cm}^{-3}$, as shown in [Fig. S5, Supporting Material]. Therefore, there is a peak in the $\zeta_{\mathbf{q}}$ in the vicinity of the K-point. Although $\zeta_{\mathbf{q}}$ of $n$-type AlN is much larger than that of $n$-type 2H-SiC, the phonon linewidth ($\Gamma_{\mathrm{pe}}=1 / 2 \tau_{\lambda}^{\mathrm{ph}-\mathrm{el}}$) of $n$-type AlN is significantly smaller than that of $n$-type 2H-SiC [Fig. S6, Supporting Material]. Therefore, the stronger EPI effects on ${\kappa}_{\text{lat}}$ of 2H-SiC cannot be interpreted by $\zeta_{\mathbf{q}}$.

\begin{figure}[H]
    \centering
    \includegraphics[width=0.70\columnwidth]{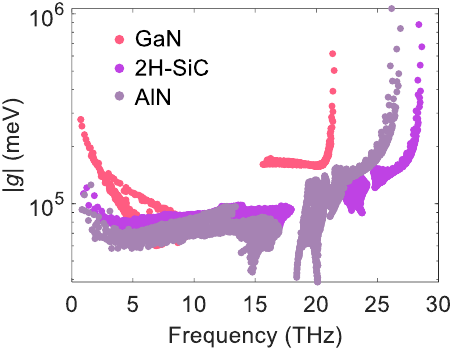}
    \caption {Absolute value of electron-phonon matrix elements $|\textit{g}|$ with respect to the phonon frequency of GaN, 2H-SiC, and AlN.}
    \label{fig: Figure 5}
\end{figure}

Note that the phonon-electron scattering rate is also related to the electron-phonon coupling strength $\textit{g}$. The electron-phonon coupling strength $\textit{g}$ quantifies the coupling strength between phonon modes and electron states,
\begin{equation}
    g_{m n}^{v}(\mathbf{k}, \mathbf{q})=\sqrt{\frac{\hbar}{2 \omega_{\lambda}}}\left\langle\psi_{m \mathbf{k}+\mathbf{q}}\left|\partial_{\lambda} V\right| \psi_{n \mathbf{k}}\right\rangle,
    \label{SE6}
\end{equation}
with $\psi$ the ground-state Bloch wave function and $\partial_{\lambda} V$ the first-order derivative of the Kohn-Sham potential with respect to the atomic displacement. The phonon-electron scattering rate has a positive relationship with $\zeta_{\mathbf{q}}$ and $|\textit{g}|$. As shown in Fig. \ref{fig: Figure 5}, the magnitude of $|\textit{g}|$ for 2H-SiC is larger than that of AlN, resulting in the smaller ${\kappa}_{\text{lat}}$ for 2H-SiC. Furthermore, although GaN has the largest magnitude of $|\textit{g}|$, its much smaller $\zeta_{\mathbf{q}}$ leads to weaker effects of EPI on ${\kappa}_{\text{lat}}$.

\section{CONCLUSIONS}
In summary, we perform a first-principles investigation on the mode-dependence phonon thermal transport in typical direct and indirect bandgap semiconductors incorporating electron-phonon interaction (EPI). It is found that EPI has a larger effect on the lattice thermal conductivity of $p$-type doping compared to $n$-type doping in the same semiconductor. The stronger EPI in $p$-type doping is attributed to the relatively higher electron density of states within the Fermi window caused by the relatively larger $p$-orbital component, which significantly increases the phonon-electron scattering channels. Moreover, EPI exhibits a stronger influence on the lattice thermal conductivity of $n$-type indirect bandgap semiconductors than $n$-type direct bandgap semiconductors. This is attributed to the relatively lower electron density of states in direct bandgap semiconductors within the Fermi window stemming from the $s$-orbital component. Nonetheless, an anomalous decline in lattice thermal conductivity is observed in the direct bandgap semiconductor AlN with $n$-type doping of $10^{21} \, \text{cm}^{-3}$, which is attributed to additional phonon-electron scattering channels created by the sub-conduction band. Moreover, it is the electron-phonon matrix element, rather than the Fermi surface nesting function, that is ascribed to the relatively larger reduction in the lattice thermal conductivity of $n$-type 2H-SiC compared to $n$-type AlN at ultrahigh concentrations.

S.L. was supported by the National Natural Science Foundation of China (Grant No. 12304039), the Shanghai Municipal Natural Science Foundation (Grant No. 22YF1400100), the Fundamental Research Funds for the Central Universities (Grant No. 2232022D-22), and the startup funding for youth faculty from the College of Mechanical Engineering of Donghua University. X.L. was supported by the Shanghai Municipal Natural Science Foundation (Grant No. 21TS1401500) and the National Natural Science Foundation of China (Grants No. 52150610495 and No. 12374027). The computational resources utilized in this research were provided by National Supercomputing Center in Shenzhen.

\bibliography{bibliography}

\end{document}